\documentclass[aps,pra,amssymb,amsmath,eqsecnum,nofootinbib]{revtex4}
\usepackage{graphicx}

\newtheorem{definition}{Definition}[section]
\newtheorem{theorem}{Theorem}[section]
\newtheorem{proposition}{Proposition}[section]

\newtheorem{remark}{Remark}[section]

\newenvironment{hypothesis}{HP: \begin{center}} {\end{center}}
\newenvironment{thesis}{TH: \begin{center}} {\end{center}}

\newtheorem{example}{Example}[section]
\begin{document}
\title{Analytic Mechanics of Locally Conservative Physical Systems}
\author{Gavriel Segre}
\homepage{http://www.gavrielsegre.com}
\begin{abstract}
 The analysis of the dynamics of a material point perfectly constrained to a submanifold of the three-dimensional euclidean space  and subjected to a locally conservative force's field, namely a force's field
 corresponding to a  closed but not necessarily exact differential
 form on such a submanifold, requires a generalization of the Lagrangian and the Hamiltonian
 formalism that is here developed.
\end{abstract}
\maketitle
\newpage
\tableofcontents
\newpage
\section{Introduction} \label{sec:Introduction}
In every elementary book of basic physics one finds the definition
of a conservative force's field as a force's field $ \vec{f} $
such that there exists a smooth function V (called energy
potential for $ \vec{f}$) such that:
\begin{equation} \label{def:first definition}
    \vec{f} \; = \; - \vec{\nabla} V
\end{equation}
Often one finds therein the statement according to which the
conservativity of a force's field is equivalent to the condition:
\begin{equation} \label{def:second definition}
   \vec{\nabla} \wedge \vec{f} \; = \; 0
\end{equation}

Following  the strategy  of converting the language of vector
calculus in the language of differential forms summarized in the
section \ref{sec:Passing from the language of vector calculus to
the language of differential forms} the equation \ref{def:first
definition} may be stated as the condition that the 1-form $
\vec{f} ^{\flat} $ is exact while the equation \ref{def:second
definition} may be stated as the condition that the 1-form $
\vec{f} ^{\flat} $ is closed.

 So one realizes that the fact that in the ordinary three
 dimensional euclidean space
 $  \mathbb{E}^{3} := ( \mathbb{R}^{3} , \delta =  dx \otimes dx + d
y \otimes d y +  dz \otimes dz  ) $  the  equation \ref{def:first
definition} and the equation \ref{def:second definition} are
indeed equivalent is a consequence of the topological triviality
of $ \mathbb{R}^{3} $: specifically of the fact that the $ 1^{th}
$ de-Rham cohomology group of $ \mathbb{R}^{3} $ is trivial (and
hence every closed form is exact, i.e. it can be globally
integrated).

Considering the dynamics of a  material point perfectly
constrained to move on a topologically non-trivial submanifold of
the euclidean space $ \mathbb{E}^{3} $ (specifically a submanifold
with non-trivial  $ 1^{th} $ de-Rham cohomology group) one
realizes that the condition \ref{def:second definition} is a
necessary but not sufficient condition for the conservativity of
the involved force's field $ \vec{f} $.

Since a force's field  satisfying the  condition of equation
\ref{def:second definition} but violating the condition
\ref{def:first definition} corresponds to a 1-form $ \vec{f}
^{\flat} $ that is closed but not exact and hence, according to
Poincar\'{e} Lemma \cite{Nakahara-03}, may be integrated locally
but not globally, it is natural to call such a force's field (and
the corresponding physical system too) \emph{locally
conservative}.

Now nonconservative force's fields performs a very short
appearance in almost all the manuals of Classical Analytical
Mechanics (see for instance \cite{Arnold-89},
\cite{Goldstein-Poole-Safko-02}) whose attention rapidly converges
on conservative systems for which the development of the
lagrangian and hamiltonian formalism is reserved.

One could, at this point, argue that, from a a physical viewpoint,
the reason for that is simple and it is incisively expressed by
Richard Feynman in the section 14.4 of \cite{Feynman-63a}:
\begin{quote}
 "We have spent a considerable time discussing conservative
 forces; what about nonconservative forces ? We shall take a
 deeper view of this than usual, and state that there are no
 nonconservative forces! As a matter of fact, all the fundamental
 forces in nature appear to be conservative. This is not a
 consequence of Netwon's Law. In fact, so far as Newton himself
 knew, the forces could be nonconservative, as friction apparently
 is. When we say friction \emph{apparently} is, we are taking a
 modern view, in which it has been discovered that all the deep
 forces, the forces between the particles at the most fundamental
 level, are conservative"
\end{quote}

As to Analytic Mechanics (defined as the mathematical discipline
dedicated to develop more advanced techniques through which
Classical Newtonian Mechanics is formalized), anyway,
nonconservative forces have to be taken into account.

They are indeed an experimental evidence of the Classical
Newtonian Physics ruling our ordinary life (being an excellent
approximation of Quantum Mechanics for macroscopic bodies as well
as an excellent approximation of Special Relativity for velocities
very much smaller than the velocity of light) and may be defined
operatively trough dynamometers.

It should be superfluous to remind that as to dissipative physical
systems (in which a portion of mechanical energy is converted into
heat), the Conservation of Energy, lost in terms of the
time-invariance of a suitable hamiltonian, is anyway guaranteed by
the First Principle of Thermodynamics.

\bigskip

In  \cite{Abraham-Marsden-78}, \cite{Marsden-Ratiu-99},
\cite{De-Azcarrega-Izquierdo-95}, \cite{Fomenko-95},
\cite{McDuff-Salamon-98}, \cite{Arnold-Givental-01} the definition
of a locally-hamiltonian vector field  $ X \in \Gamma ( T M  ) $
over a symplectic manifold $ ( M , \omega ) $ as a vector field
such that the one-form $ i_{X} \omega $ is closed (but not
necessarily exact) is presented.

No systematic analysis about how to develop the hamiltonian
formalism for locally-hamiltonian vector fields is, anyway,
developed therein.

Clearly, from a mathematical viewpoint, locally conservative
physical systems are indeed a particular case of locally
hamiltonian vector fields.

From a physical point of view, anyway, considering locally
conservative physical systems in such a general framework doesn't
allow to focalize the attention to the basic physical entity: the
underlying force's field.

For this reason we have chosen to restrict our analysis to
locally-conservative physical systems, whose analysis requires an
extension of the lagrangian and hamiltonian formalism presented in
this paper \footnote{As to the following quotation by Vladimir
Arnold \cite{Arnold-2005}:
\begin{quote}
 "There are two principal ways to formulate mathematical
 assertions (problems, conjectures, theorems, $ \cdots $): Russian
 and French. The \emph{Russian way} is to choose \emph{the most simple and specific
 case} (so that nobody could simplify the formulation preserving
 the main point). The \emph{French way} is \emph{to generalize the statement as far as nobody could generalize it further"}
\end{quote}
the french approach to this paper would involve sheaf cohomology,
Grothendieck topologies and much more; the russian approach would
restrict the paper to  the example \ref{ex:prototype1},  example
\ref{ex:prototype2}, example \ref{ex:prototype3}, example
\ref{ex:prototype4} and example \ref{ex:prototype5}. As an italian
jew I have chosen an intermediate approach.}.

\smallskip

The issue discussed in this paper is clearly related to the
extension of Morse Theory to multivalued functions exposed in the
first appendix of \cite{Dubrovin-Novikov-Fomenko-92}.

We won't follow  this conceptual path, observing that no
systematic foundation of Mechanics based on some Principle of
Minimal Action for multivalued action functionals is therein
presented.

\smallskip

We will analyze, in particular, the paradigmatic example of the
locally conservative force's field $ \frac{- y}{x^{2} + y^{2} } dx
+ \frac{x}{x^{2} + y^{2} } d y $ on the punctured plane $
\mathbb{R}^{2} - \{ \vec{0} \} $.

\newpage
\section{Set of local energy potentials for locally conservative force's fields}
Let us suppose to have a physical system consisting of a material
point of mass $ m \in ( 0 , + \infty ) $  perfectly constrained to
a submanifold \footnote{To lighten the terminology we will denote
from this time forward simply with the term \emph{manifold } an
\emph{arcwise connected differentiable manifold}.} M of the
3-dimensional euclidean space $ \mathbb{E}^{3} := ( \mathbb{R}^{3}
, \delta = dx \otimes dx + d y \otimes d y + dz \otimes dz  ) $
subjected to the force's field $ f \in \Omega^{1} ( M) $
\footnote{where we have implicitly followed the strategy of
converting the language of vector calculus in the language of
differential forms summarized in the section \ref{sec:Passing from
the language of vector calculus to the language of differential
forms} by assuming from this time forward that the Flat operation
$ \flat $ is undertstood.}, where, following the terminology and
the notation of \cite{Nakahara-03}, $ \Omega^{r} ( M) $ is the set
of the r-forms over M, $ Z^{r} (M) $ is the \emph{ $ r^{th} $
cocycle group} of M, $ B^{r} (M) $ is the \emph{$ r^{th} $
coboundary group of M} and $ H^{r} (M) := \frac{Z^{r} (M)}{B^{r}
(M)} $ is the \emph{$r^{th}$ de Rham cohomology group} of M
\footnote{In general, given a group G, one defines the \emph{$
n^{th}$ cohomology group of M with respect to the group G} as
\cite{Nash-94} :
\begin{equation}
    H^{n} (M ; G ) \; := \; [ M , K(G,n) ]_{0}
\end{equation}
where $ [ X , Y ]_{0} $ denotes the set of based homotopy classes
of $ Y^{X} $ and where $ K(G, n ) $ denotes the Eilenberg- Mac
Lane spaces defined, up to homotopic equivalence, by the
condition:
\begin{equation}
    \pi_{m} ( K(G,n)) \; = \;
    \begin{cases}
    G, & \hbox{if $ m = n$;} \\
    \{ \mathbb{I} \} , & \hbox{otherwise.} \\
\end{cases}
\end{equation}
Such a notion may be further generalized defining, in a suitable
way, the groups of cohomology $ H^{n}( M , \mathcal{S} ) $ of M
with respect to a sheaf (see appendix \ref{sec:The double meaning
of the locution "Riemann Surface"}). Following the terminology of
\cite{Nakahara-03} we denote the $ n^{th} $ group of De Rham
cohomology of M as:
\begin{equation}
    H^{n}( M ) \; = \; H^{n} ( M , \mathbb{R} ) \; = \; H^{n} ( M ,
    \mathcal{S}_{constant})
\end{equation}
 (where $ \mathcal{S}_{constant} $ is the sheaf of constant functions over
M). As to homology groups let us observe that, under mild
topological conditions (see the section 6.1.1 of
\cite{Nakahara-03} and the section 1.6 of
\cite{Dubrovin-Novikov-Fomenko-92}) that we will assume from this
time foward, $ H_{n}( M ; \mathbb{Z} )$ and $ H_{n}( M ;
\mathbb{R} )$ are isomorphic; hence we will simply talk about the
$ n^{th} $ homology group of M:
 \begin{equation}
    H_{n} ( M) \; = \; H_{n} ( M , \mathbb{R} )
\end{equation}
}.

Let us recall that:
\begin{definition}
\end{definition}
\emph{f is conservative:}
\begin{equation}
    f \in B^{1} (M)
\end{equation}

\smallskip

We will say that:

\begin{definition}
\end{definition}
\emph{f is locally conservative:}
\begin{equation*}
    f \in Z^{1} (M)
\end{equation*}

We will refer to the material point of mass $ m \in ( 0 , + \infty
)$ perfectly constrained to move on  M under the influence of a
locally conservative force's field as to a locally conservative
physical system.

\begin{remark}
\end{remark}
If $ H^{1} ( M ) = \{ \mathbb{I} \} $ a locally conservative
force's field is also conservative.

We  will assume from this time forward that this is not the case
and we will restrict the analysis to the situation in which f is
locally conservative but it is not conservative.

Hence $ [ f] \in H^{1} ( M ) $ is a non-trivial cohomology class.

\bigskip

\begin{remark}
\end{remark}
A physical system whose  mathematical structure is similar (but
different) to the one discussed in this paper consists of a
material point of mass $ m \in ( 0 , + \infty ) $ and electric
charge $ e \in \mathbb{R} - \{ 0 \} $ perfectly constrained to
move on a submanifold M of the 3-dimensional euclidean space $
\mathbb{E}^{3} := ( \mathbb{R}^{3} , \delta := dx \otimes dx + d y
\otimes d y +  dz \otimes dz     ) $ under the influence of the
Lorentz's force' $ f = \frac{e}{c} v \wedge B $ (where $ v = v_{x}
dx + v_{y} dy + v_{z} dz  \in \Omega^{1}(M) $ is the velocity of
the material point while $ B = B_{x} dy \wedge dz - B_{y} dx
\wedge dz + B_{z} dx \wedge dy \in Z^{2} (M) $ is the magnetic
field) induced by a magnetic field $ B \notin B^{2} (M) $ (and
hence $ [ B ] \in H^{2} ( M ; \mathbb{R}  ) $ is a not-trivial
cohomology class) extensively studied in the literature (see for
instance \cite{Balachandran-Marmo-Skagerstam-Stern-83},
\cite{De-Azcarrega-Izquierdo-95}).

\bigskip

\begin{figure}
  \includegraphics[scale=.5]{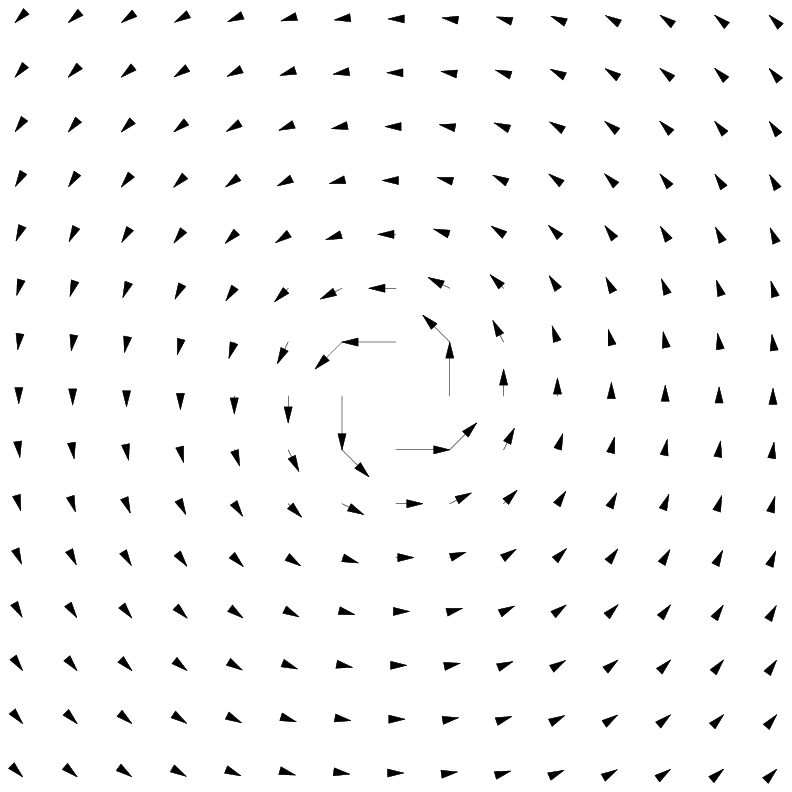}\\
  \caption{The force's field f} \label{fig.prototype}
\end{figure}

\begin{example} \label{ex:prototype1}
\end{example}
Let us suppose that $ M = \mathbb{R}^{2} - \{ \vec{0} \} $ and let
us introduce the following  force's field:
\begin{equation}
    f := f_{x}  dx \, + \, f_{y}
    dy \in \Omega^{1} (M)
\end{equation}
\begin{eqnarray}
  f_{x} &:=& \frac{- y}{ x^{2} + y^{2} } \\
   f_{y} &:=& \frac{x}{ x^{2} + y^{2} }
\end{eqnarray}
represented in the figure \ref{fig.prototype}.

Since:
\begin{equation}
    \frac{\partial f_{x}}{ \partial y } \; = \; \frac{y^{2} - x^{2} }{( x^{2} +
    y^{2})^{2}} \; = \;  \frac{\partial f_{y}}{ \partial x }
\end{equation}
it follows that:
\begin{equation}
    d f \; = \; ( \frac{\partial f_{y}}{ \partial x } -  \frac{\partial f_{x}}{ \partial y }
    ) d x \wedge dy \; = \; 0
\end{equation}
and hence $ f \in Z^{1}(M) $.

Let us now introduce the map $ \theta : M - \{ ( 0 ,y) , y \in ( -
\infty , 0) \cup ( 0 , + \infty ) \} \mapsto \mathbb{R} $:
\begin{equation}
    \theta ( x , y) \; := \; \arctan ( \frac{y}{x} )
\end{equation}
One may easily verify that:
\begin{equation}
    d \theta \; := \frac{\partial \theta}{\partial x } dx \, + \,\frac{\partial \theta}{\partial y } dy \; = \;  f_{| M - \{ ( 0 ,y) , y \in ( -
\infty , 0) \cup ( 0 , + \infty ) \} }
\end{equation}
where $ f_{| M - \{ ( 0 ,y) , y \in ( - \infty , 0) \cup ( 0 , +
\infty ) \}} \in \Omega^{1} (   M - \{ ( 0 ,y) , y \in ( - \infty
, 0) \cup ( 0 , + \infty ) \}    )  $ is the restriction of f to $
M - \{ ( 0 ,y) , y \in ( - \infty , 0) \cup ( 0 , + \infty ) \} $.

It follows that:
\begin{enumerate}
    \item
\begin{equation}
   f_{| M - \{ ( 0 ,y) , y \in ( - \infty , 0) \cup ( 0 , +
\infty ) \}} \in B^{1} ( M - \{ ( 0 ,y) , y \in ( - \infty , 0)
\cup ( 0 , + \infty ) \}  )
\end{equation}
    \item
\begin{equation}
    f \; \notin \; B^{1} ( M )
\end{equation}
\end{enumerate}

\bigskip

\begin{remark}
\end{remark}
The closed but not exact 1-form of the example \ref{ex:prototype1}
has been extensively studied in a different physical context
\cite{Lerda-92} as the vector-potential 1-form on the punctured
plane x,y $ \mathbb{R}^{2} - \{ \vec{0} \} $ generated by a
 magnetic flux line along the z-axis.

The more famous physical appearance of the 1-form $
\frac{-y}{x^{2}+y^{2}} dx + \frac{x}{x^{2}+y^{2}} dy $ is in the
context of the hydrodynamical analogue of such a system (where
hence such a form represents an Eulerian velocity's field) as a
vortex.

\bigskip

Following once more the terminology and the notation of
\cite{Nakahara-03} let  $ C_{r} (M) $ be the \emph{$ r^{th} $
chain group of M,} let $ Z_{r} (M) $ be the \emph{$ r^{th}$ cycle
group of M}, let $ B_{r} (M) $ be the \emph{$ r^{th}$ boundary
group of M}, let $ H_{r} (M) := \frac{ Z_{r} (M) }{B_{r} (M) } $
be the \emph{$ r^{th}$ homology group of M.} and let $ \pi_{r} (M)
$ be the \emph{$ r^{th} $ homotopy group} of M.

Let us recall that given a field force $ f \in \Omega^{1} (M) $
and a 1-cycle $ c \in C_{1} (M) $:

\begin{definition}
\end{definition}
\emph{work made by f along c:}
\begin{equation}
    W( c , f)  \; := \; \int_{c} f
\end{equation}

Given a conservative field force $ f \in B^{1} (M) $:
\begin{definition}
\end{definition}
\emph{potential energy of f:}
\begin{equation}
    V \in \Omega^{0} (M) \; : \; f = - d V
\end{equation}

\bigskip

\begin{remark}
\end{remark}
Let us remark that f is invariant under the following action of $
\mathbb{R} $ (seen as an abelian group) over $ \Omega^{0} (M) $:
\begin{equation}
    a \in \mathbb{R} : V \mapsto V + a
\end{equation}
in the following sense: if V is an energy potential of f then $
V+a $ is also an energy potential of f  for every $ a \in
\mathbb{R} $.

\bigskip

Applying Stokes' Theorem the work made by a conservative force's
field with energy potential V along the 1-cycle $ c \in C_{1} (M)
$ may be written as:
\begin{equation}
  W( c , f) \; = \; - W( c ,  d V ) \; = \; - W( \partial c , V)
\end{equation}

So the work made by a conservative force's field with energy
potential V along a 1-cycle c depends only by the values taken by
V on the boundary of c.

Let us now recall the Hurewicz's isomorphism:
\begin{equation} \label{eq:Hurewicz's isomorphism}
    H_{1}(M) \; = \frac{\pi_{1}(M) }{[ \pi_{1} (M) ,  \pi_{1} (M) ] }
\end{equation}
where we have denoted by [ G , G ] the commutator subgroup of an
arbitary group G defined as:
\begin{equation}
  [ G , G ] \; := \; \{ x \cdot y \cdot x^{-1} \cdot y^{-1} \; \; x,y \in G  \}
\end{equation}
It follows that if $ \pi_{1} (M) $ is abelian then $ H_{1} ( M) =
\pi_{1} (M)$.

\bigskip

\begin{remark}
\end{remark}
Let us remark, by the way, that equation \ref{eq:Hurewicz's
isomorphism} implies that a locally conservative but not
conservative force field may exist only on a multiply connected
manifold.

It is curious, with this regard, that though the formalism of
Quantum Mechanics on multiply connected configuration spaces,
pioneered at the end of the  sixthes and the begininning of the
seventhes by Larry Schulman and Cecile Morette De Witt, is
nowadays commonly founded in  the literature (see for instance the
$ 23^{th}$ chapter of \cite{Schulman-81}, the $ 7^{th} $ chapter
of \cite{Rivers-87} and the $ 8^{th} $ chapter of
\cite{Cartier-De-Witt-Morette-06} as to its implementation, at
different levels of mathematical rigor, in the path-integration's
formulation, as well as the $ 8^{th} $ chapter of
\cite{Balachandran-Marmo-Skagerstam-Stern-91}, the $ 3^{th} $
chapter of \cite{Morandi-92} and the section 6.8 of
\cite{Strocchi-05b} for its formulation in the operatorial
formulation) \footnote{in presence of a multiply connected
configuration space a suppletive topological superselection rule
exists, the involved superselection charge, taking values on $
Hom( H_{1}(\text{configuration space}) , U(1))$, appearing in
physically very different contexts going from the $ \theta$-angle
of Yang-Mills quantum field theories (see for instance the
$10^{th} $ chapter of \cite{Ryder-96} or the section 23.6 of
\cite{Weinberg-96}) to the magnetic flux of the solenoid involved
in the Aharonov-Bohm effect and to the fractional statistic of a
quantum system of identical particles living on the plane (see for
instance \cite{Wilczek-90} and \cite{Lerda-92})}, the role of the
multiple-connectivity of the configuration space in Classical
Mechanics is, at least up to our knowledge, largely unexplored.

For instance, up to our knowledge, no systematic comparison of the
\emph{left generalized rigid body} of a Lie group G (defined,
according to the $2^{th}$ appendix of \cite{Arnold-89}, as the
dynamical system with lagrangian $ L : T G \mapsto \mathbb{R} $:
\begin{equation} \label{eq:lagrangian of a rigid body}
    L( q, \dot{q} ) \; := \; \frac{1}{2} | \dot{q} |_{g}^{2}
\end{equation}
where g is the left-invariant riemannian metric on G) and the
\emph{left generalized rigid body} of $ \tilde{G} $ (where $
\tilde{G}$ is the universal covering group of G) exists in the
literature, though its importance, for instance, in the cases $ G
:= SO(n) $, $ \tilde{G} =: Spin(n) $.

The same can be said as to the comparison between the \emph{right
generalized rigid body} of a Lie group G (defined as the dynamical
system with lagrangian $ L : T G \mapsto \mathbb{R} $ given by the
equation \ref{eq:lagrangian of a rigid body} where g is the
right-invariant riemannian metric of G)  and the \emph{right
generalized rigid body} of $ \tilde{G} $ though its importance,
for instance, in Fluid Dynamics, where G is the group of the
diffeomorphisms of the manifold on which the fluid moves, and in
the related Physics of solitons  (considering for example that the
Korteweg de Vries equation may be interpreted as the motion's
equation of the right generalized rigid body of the Virasoro group
$\widetilde{ Diff( S^{1})} $ \cite{Marsden-Ratiu-99}).

\bigskip

Let us assume that  $ \pi_{1} (M) $ is abelian.

 Given a path $ \alpha : [ 0 ,
1 ] \mapsto M $ one has then that:
\begin{equation}
    W( \alpha , f ) \; = \; V( \alpha(1)) -  V( \alpha(0))
\end{equation}

It follows that the work made by a conservative force's field
along a loop is equal to zero.

The work made by a locally conservative force's field along a loop
may, contrary, be different from zero.

\begin{example} \label{ex:prototype2}
\end{example}
In the framework of the example \ref{ex:prototype1} let us observe
first of all that the fundamental group of the punctured plane is
abelian:
\begin{equation}
    \pi_{1} ( M ) \; =  \; \mathbb{Z}
\end{equation}
and hence:
\begin{equation}
    H_{1}( M ) \; = \; \pi_{1} ( M ) \; =  \; \mathbb{Z}
\end{equation}

Given a loop $ c : [ 0 , 1 ] \mapsto M $ it may be easily verified
that:
\begin{equation} \label{eq:work along a loop}
    W( f , c ) \; = \; 2 \pi n_{winding} ( c , \vec{0} )
\end{equation}
where in general $ n_{winding} ( c , q) $ is the winding number of
the loop c with respect to the point q.

\bigskip

Let us now consider a locally conservative field force $ f \in
Z^{1} (M) $.

Let us recall that by Poincar\'{e} Lemma it follows that for every
contractible open set  U there exists a $ V_{U} \in \Omega^{0} ( U
) $ such that:
\begin{equation}
    f_{| U} \; = - d V_{U}
\end{equation}
where $  f_{| U} \in B^{1}(U) $ is the restriction of f to the
open set U.

This allows to introduce the following:
\begin{definition} \label{def:set of local energy potentials of a locally conservative force's field}
\end{definition}
\emph{set of local energy potentials for f:}

a set $ \{ V_{i}  \}_{i=1}^{n} $ such that:
\begin{enumerate}
  \item
\begin{equation}
    n \in \mathbb{N}_{+}
\end{equation}
  \item
\begin{equation}
   V_{i} \; \in \; \Omega^{0} ( U_{i} ) \; \; i = 1 , \cdots , n
\end{equation}
    \item $ \{ U_{i} \}_{i=1}^{n} $ is a covering of M such that:
\begin{equation}
    U_{i}  \text{ is contractible } \; \; \forall i \in \{ 1 ,
    \cdots , n \}
\end{equation}
    \item
\begin{equation}
    f_{| U_{i}} \; = - d V_{i} \; \; \forall i \in \{ 1 ,
    \cdots , n \}
\end{equation}
\end{enumerate}

\bigskip

\begin{remark}
\end{remark}
Given a locally conservative force's field f and a set $ \{ V_{i}
\}_{i=1}^{n} $ of local energy potentials for f, let us observe
that definition \ref{def:set of local energy potentials of a
locally conservative force's field} implies that if $ U_{i} \cap
U_{j} \; \neq \; \emptyset $, then the map $ c_{ij} : U_{i} \cap
U_{j} \mapsto \mathbb{R} $ defined by:
\begin{equation}
  c_{ij} (q) \; := \;   V_{i}(q) - V_{j}(q)
\end{equation}
is constant.

Let us observe furthermore that:
\begin{enumerate}
    \item
\begin{equation} \label{eq:first condition}
    c_{ii} \; = \; 0
\end{equation}
    \item
\begin{equation} \label{eq:second condition}
    c_{ij} \; = \; - c_{ji}
\end{equation}
    \item if $ U_{i} \cap
U_{j} \cap U_{k}  \; \neq \; \emptyset $ then:
\begin{equation} \label{eq:third condition}
    c_{ij} + c_{jk} \; = c_{ik}
\end{equation}
\end{enumerate}

\bigskip

\begin{remark}
\end{remark}
Let us remark that f is invariant under the following action of $
\mathbb{R}^{n} $ (seen as an abelian group):
\begin{equation}
    (a_{1}, \cdots , a_{n} ) \in \mathbb{R}^{n} \; : \; V_{i}
    \mapsto V_{i} + a_{i} \; \; \forall i \in \{ 1 , \cdots , n \}
\end{equation}
in the following sense: if  $ \{ V_{i} \}_{i=1}^{n} $ is a set of
local energy potentials of the locally conservative force's field
f then $  \{ V_{i} + a_{i} \}_{i=1}^{n} $ is also a set of local
energy potentials for f.

Clearly:
\begin{equation}
    (a_{1}, \cdots , a_{n} ) \in \mathbb{R}^{n} \; : \; c_{ij}
    \mapsto c_{ij} + a_{i} - a_{j} \; \; \forall i,j \in \{ 1 , \cdots , n \}
\end{equation}

\bigskip

\begin{example} \label{ex:prototype3}
\end{example}
In the framework of the example \ref{ex:prototype1} and of the
example \ref{ex:prototype2}  let us observe first of all that:
\begin{equation}
    \lim_{x \rightarrow 0^{+}} \theta (x,y) \; = \; +
    \frac{\pi}{2} \; \; \forall y \in ( 0 , + \infty )
\end{equation}
\begin{equation}
    \lim_{x \rightarrow 0^{-}} \theta (x,y) \; = \; -
    \frac{\pi}{2} \; \; \forall y \in ( 0 , + \infty )
\end{equation}
\begin{equation}
    \lim_{x \rightarrow 0^{+}} \theta (x,y) \; = \; -
    \frac{\pi}{2} \; \; \forall y \in ( - \infty , 0 )
\end{equation}
\begin{equation}
    \lim_{x \rightarrow 0^{-}} \theta (x,y) \; = \; +
    \frac{\pi}{2} \; \; \forall y \in ( - \infty , 0 )
\end{equation}
and hence in particular:
\begin{equation}
    \nexists \; \lim_{x \rightarrow 0} \theta (x,y) \; \; \forall
    y \in ( - \infty , 0 ) \cup ( 0 , + \infty )
\end{equation}
so that it doesn't exist a continuous extension of the map $
\theta $ to the whole M.

Let us consider the following  covering $ \{ U_{i} \}_{i=1}^{4}  $
of M:
\begin{equation}
    U_{1} \; := \; \{ ( x,y) \in \mathbb{R}^{2} \; : \; x \geq 0 \wedge y \geq 0
    \} - \{ \vec{0} \}
\end{equation}
\begin{equation}
  U_{2} \; := \; \{ ( x,y) \in \mathbb{R}^{2} \; : \; x \leq 0 \wedge y \geq 0
    \} - \{ \vec{0} \}
\end{equation}
\begin{equation}
   U_{3} \; := \; \{ ( x,y) \in \mathbb{R}^{2} \; : \; x \leq 0 \wedge y \leq 0 \} - \{ \vec{0} \}
\end{equation}
\begin{equation}
    U_{4} \; := \; \{ ( x,y) \in \mathbb{R}^{2} \; : \; x \geq 0 \wedge y \leq 0
    \} - \{ \vec{0} \}
\end{equation}
Clearly:
\begin{equation}
    U_{i} \text{ is contractible } \; \; \forall i \in \{ 1 , \cdots , 4 \}
\end{equation}
Let us the introduce the following maps:
\begin{enumerate}
    \item $ V_{1} \in \Omega^{0} ( U_{1} ) $:
\begin{equation}
    V_{1} ( x, y ) \; := \; \left\{%
\begin{array}{ll}
    - \theta ( x , y), & \hbox{if $ x > 0$;} \\
    + \frac{\pi}{2}, & \hbox{if $ x = 0 $.} \\
\end{array}%
\right.
\end{equation}
    \item  $ V_{2} \in \Omega^{0} ( U_{2} ) $:
\begin{equation}
    V_{2} ( x, y ) \; := \; \left\{%
\begin{array}{ll}
    - \theta ( x , y), & \hbox{if $ x < 0$;} \\
    -  \frac{\pi}{2}, & \hbox{if $ x = 0 $.} \\
\end{array}%
\right.
\end{equation}
    \item $ V_{3} \in \Omega^{0} ( U_{3} ) $:
\begin{equation}
    V_{3} ( x, y ) \; := \; \left\{%
\begin{array}{ll}
    - \theta ( x , y), & \hbox{if $ x < 0$;} \\
    +  \frac{\pi}{2}, & \hbox{if $ x = 0 $.} \\
\end{array}%
\right.
\end{equation}
    \item $ V_{4} \in \Omega^{0} ( U_{4} ) $:
\begin{equation}
    V_{4} ( x, y ) \; := \; \left\{%
\begin{array}{ll}
    - \theta ( x , y), & \hbox{if $ x > 0$;} \\
    -  \frac{\pi}{2}, & \hbox{if $ x = 0 $.} \\
\end{array}%
\right.
\end{equation}
\end{enumerate}

By construction we have that $ \{ V_{i} \}_{i=1}^{4} $ is a set of
local energy potentials for f.

Clearly:
\begin{equation}
    c_{12} \; = \; + \pi
\end{equation}
\begin{equation}
    c_{23} \; = \; - \pi
\end{equation}
\begin{equation}
    c_{34} \; = \; + \pi
\end{equation}
\begin{equation}
    c_{41} \; = \; - \pi
\end{equation}
while, since $ U_{1} \cap U_{3} \; = \; U_{2} \cap U_{4} \; = \;
\emptyset $, $c_{13} $ and $ c_{24} $ are undefined.
\newpage
\section{The generalization of  the Lagrangian and the Hamiltonian formalism required to analyze locally conservative physical systems}
Given the locally conservative (but not conservative) physical
system (that we will denote as $ PS_{f} $) consisting of a
material point of mass $ m \in ( 0 , + \infty ) $ perfectly
constrained to a differentiable submanifold M of the 3-dimensional
euclidean space $ \mathbb{E}^{3} := ( \mathbb{R}^{3} , \delta = dx
\otimes dx + d y \otimes d y +  dz \otimes dz  ) $ subjected to
the locally conservative (but not conservative) force's field f we
will show how its analysis in the framework of Analytical
Mechanics requires the introduction of a suitable generalization
of both the Lagrangian and the Hamiltonian formalism.

Let us observe first of all that (assuming that $ \{ \vec{0} , \{
\vec{e_{1}} , \vec{e}_{2}, \vec{e}_{3} \} \} $ is an inertial
frame) the dynamics of $ PS_{f} $ is ruled by the Newton's Law:
\begin{equation}
    m \ddot{\vec{r}} \; = \; \vec{f} + \vec{f}_{constraint}
\end{equation}
where $ \vec{f}_{constraint} $ is the force's field constraining
the material point to M.

\bigskip

\begin{remark}
\end{remark}
Let us remark that we define a perfect constraint as one
satisfying the following \emph{Generalized D'Alambert - Lagrange's
Principle}:
\begin{equation} \label{eq:generalized D'Alambert Lagrange's Principle}
    W ( f_{constraint} , c ) \; = \; 0 \; \; \forall c \in
    C_{1}(M)
\end{equation}

that, if $ \pi_{1} (M) $ is abelian, (owing to equation
\ref{eq:Hurewicz's isomorphism}) reduces to the \emph{(Ordinary)
D'Alambert - Lagrange's Principle}:
\begin{equation}\label{eq:ordinary D'Alambert Lagrange's Principle}
    W ( f_{constraint} , \alpha ) \; = \; 0 \; \; \forall  \text{ path $ \alpha $ on M}
\end{equation}

 Let us now introduce the following:
\begin{definition}
\end{definition}
\emph{set of local lagrangians of $ PS_{f} $:}

a set $ \{ L_{i} \}_{i=1}^{n} $ such that $  L_{i} : T U_{i}
\mapsto \mathbb{R} $ is defined as:
\begin{equation}
    L_{i} ( q , \dot{q} ) \; := \; \frac{m}{2} | \dot{q} |_{g}^{2} -
    V_{i} (q)
\end{equation}
where $  \{ V_{i} \}_{i=1}^{n} $ is a set of local energy
potentials for f and where $ g := i^{\star} \delta $ is the
riemannian metric over M induced by the inclusion's embedding $ i
: M \mapsto \mathbb{R}^{3} $.

\begin{definition} \label{def:local hamiltonians}
\end{definition}
\emph{set of local hamiltonians of $ PS_{f} $:}

a set $ \{ H_{i} \}_{i=1}^{n} $ such that $  L_{i} : T^{\star}
U_{i} \mapsto \mathbb{R} $ is defined as:
\begin{equation}
    H_{i} (p , q) \; := \; \frac{|p|_{g}^{2}}{2 m} + V_{i} ( q)
\end{equation}
where $  \{ V_{i} \}_{i=1}^{n} $ is a set of local energy
potentials for f and where $ g := i^{\star} \delta $ is the
riemannian metric over M induced by the inclusion's embedding $ i
: M \mapsto \mathbb{R}^{3} $.

\bigskip

\begin{remark}
\end{remark}
Let us remark that each local lagrangian $ L_{i} $ of a set of
local lagrangians $ \{ L_{i} \}_{i=1}^{n} $ for $ PS_{f} $ is
regular.

Each local hamiltonian $ H_{i} $ of the set of local hamiltonians
$ \{ H_{i} \}_{i=1}^{n} $ for $ PS_{f} $ constructed with the same
set of local energy potentials of $ \{ L_{i} \}_{i=1}^{n} $ can
then be simply obtained from $ L_{i} $ through Legendre's
transform.

\bigskip

Given $ i \in \{ 1 , \cdots , n \} $:
\begin{definition}
\end{definition}
\emph{dynamical system $ DS_{i} $:}
\begin{center}
 the dynamical system having lagrangian $ L_{i} $ (and hence having
 hamiltonian $ H_{i} $)
\end{center}

\bigskip

\begin{remark}
\end{remark}
To each dynamical system $ DS_{i} $ one can apply  the theorem of
the section 21 of the $ 4^{th} $ chapter of \cite{Arnold-89}
stating its equivalence with the (Ordinary) D'Alambert Principle
as well as its equivalence with the limit $ N \rightarrow + \infty
$ of a system of energy potential $ V_{i} + N V_{constraint} $
where:
\begin{equation}
    V_{constraint}( \vec{x} , M ) \; := \; distance_{\delta}^{2} (
    \vec{x} , M)
\end{equation}
($ distance_{\delta} (
    \vec{x} , M) $ being of course the distance, with respect of
    the euclidean metric $ \delta $, of the point $ \vec{x} $ from M).

\bigskip

\begin{remark}
\end{remark}
The definition \ref{def:local hamiltonians} allows to appreciate
how locally conservative physical systems can be seen as
particular cases of locally hamiltonian vector fields
 \cite{Abraham-Marsden-78}, \cite{Marsden-Ratiu-99}, \cite{De-Azcarrega-Izquierdo-95}, \cite{Fomenko-95}, \cite{McDuff-Salamon-98} \footnote{We advise the reader that in \cite{McDuff-Salamon-98} locally hamiltonian vector
 fields are called symplectic vector fields.}.

Given a symplectic manifold $ ( Q , \omega )$ and a vector field $
X \in \Gamma ( T Q ) $ let us recall that:
\begin{definition}
\end{definition}
\emph{X is hamiltonian:}
\begin{equation}
    i_{X} \omega \; \in \; B^{1} (Q)
\end{equation}
\begin{definition}
\end{definition}
\emph{X is locally-hamiltonian:}
\begin{equation}
    i_{X} \omega \; \in \; Z^{1} (Q)
\end{equation}
An hamiltonian vector field is obviously also a locally
hamiltonian vector field.

Given an hamiltonian vector field X, a function $ H \in \Omega^{0}
(Q) $ such that $ i_{X} \omega = d H $ is called an hamiltonian of
X.

If $ H^{1} ( Q) \neq \{ I \} $ a locally hamiltonian vector field
$ X \in \Gamma ( T Q ) $ is not in general an hamiltonian vector
field.

Let us introduce the following:
\begin{definition}
\end{definition}
\emph{Lie group of the symplectomorphisms of $ ( Q , \omega )$:}
\begin{equation}
    Simp( Q , \omega ) \; := \; \{ \psi \in Diff(Q) \, : \, \psi^{\star} \omega = 0 \}
\end{equation}

If Q is closed (i.e. compact and without boundary) then the set of
the locally-hamiltonian vector fields is the Lie algebra of the
Lie group $ Simp( Q , \omega ) $.

\smallskip

A general analysis concerning the flows of locally hamiltonian
vector fields is, up to or knowledge, still lacking.

For the reasons explained in section \ref{sec:Introduction} we
will not pursue such a general approach, limiting our attention to
locally conservative physical systems that, as will now show, are
nothing but a particular case.

Poincar\'{e} Lemma assures us  that given a covering $ \{ U_{i}
\}_{i=1}^{n} $ of Q such that:
\begin{equation}
    U_{i} \text{ is contractible } \; \; \forall i \in \{ 1 ,
    \cdots , n \}
\end{equation}
one has that:
\begin{equation}
    i_{X} \omega \; \in \; B^{1} (U_{i}) \; \;
\end{equation}
and hence there exists a set $ \{ H_{i} \}_{i=1}^{n} $ such that:
\begin{enumerate}
    \item
\begin{equation}
    H_{i} \in \Omega^{0} (U_{i} ) \; \;  \forall i \in \{ 1 ,
    \cdots , n \}
\end{equation}
    \item
\begin{equation}
   i_{X} \omega \; = \; d H_{i} \; \; \forall i \in \{ 1 ,
    \cdots , n \}
\end{equation}
\end{enumerate}

It is natural to call the set $ \{ H_{i} \}_{i=1}^{n} $ a set of
local hamiltonians for X.

Let us now consider the particular case in which the symplectic
manifold $ ( Q , \omega ) $ is of the form $ ( T^{\star} M ,
\omega_{can} ) $ for a suitable differential submanifold M of the
three dimensional euclidean space
 $  \mathbb{E}^{3} := ( \mathbb{R}^{3} , \delta =  dx \otimes dx + d
y \otimes d y +  dz \otimes dz  ) $ (where $ \omega_{can} $ is the
canonical symplectic form of the cotangent bundle $ T^{\star} M
$).

Given a locally conservative force's field $ f \in Z^{1}(M) $ one
can find a suitable locally hamiltonian vector field $ X_{f} \in
\Gamma ( T^{\star} M ) $ whose corresponding set of local
hamiltonians is a set of local hamiltonians for $ PS_{f} $ in the
sense of the definition \ref{def:local hamiltonians}.

\bigskip

A local lagrangian $ L_{i} $ or hamiltonian $ H_{i} $ can be used
to derive, through respectively the Euler-Lagrange's equation or
the Hamilton's equations, the motion  $ q( t) $ associated to an
initial condition  $ q(0) \in U_{i} $ and  such that:
\begin{equation} \label{eq:invariance constraint}
    q(t) \in U_{i} \; \; \forall t \in (0, + \infty )
\end{equation}

\bigskip

\begin{example} \label{ex:prototype4}
\end{example}
In the framework of the example \ref{ex:prototype1}, of the
example \ref{ex:prototype2} and of the example \ref{ex:prototype3}
let us observe first of all that obviously, according to Newton's
Law (assuming that $ \{ \vec{0} , \{ \vec{e_{1}} , \vec{e}_{2},
\vec{e}_{3} \} \} $ is an inertial frame), the dynamics of $
PS_{f} $ is ruled by the differential equations:
\begin{equation}
    m \ddot{x} \; = \; \frac{- y}{x^{2}+y^{2}}
\end{equation}
\begin{equation}
    m \ddot{y} \; = \; \frac{x}{x^{2}+y^{2}}
\end{equation}
\begin{equation}
    m \ddot{z} \; = \;  f_{constraint}
\end{equation}

A set of local lagrangians for  $ PS_{f} $ is given by:
\begin{equation}
    L_{i} ( x,y , \dot{x} , \dot{y} ) \; = \; \frac{m}{2} (
    \dot{x}^{2}+ \dot{y}^{2} ) - V_{i}( x,y) \; \; i \in \{ 1 , \cdots, 4 \}
\end{equation}
while a set of local hamiltonians for $ PS_{f} $ is given by:
\begin{equation}
    H_{i} ( x, y , p_{x} , p_{y} ) \; = \; \frac{ p_{x}^{2} + p_{y}^{2}
    }{2m} + V_{i} (x,y) \; \; i \in \{ 1 , \cdots, 4 \}
\end{equation}
where:
\begin{equation}
     H_{i} ( x, y , p_{x} , p_{y} ) \; = \;  p_{x} \dot{x} + p_{y}
     \dot{y} -  L_{i} ( x,y , \dot{x} , \dot{y} ) \; \; i \in \{ 1 , \cdots, 4 \}
\end{equation}
\begin{equation}
    p_{x} \; = \; \frac{ \partial L_{i} }{ \partial \dot{x} } \; =
    \; m \dot{x} \; \; i \in \{ 1 , \cdots, 4 \}
\end{equation}
\begin{equation}
    p_{y} \; = \; \frac{ \partial L_{i} }{ \partial \dot{y} } \; =
    \; m \dot{y} \; \; i \in \{ 1 , \cdots, 4 \}
\end{equation}

Let us now concentrate our attention to $ U_{1} $.

Passing to polar coordinates $ \delta = dr \otimes dr + r^{2} d
\theta \otimes d \theta $ and observing that:
\begin{equation}
    V_{1} ( r , \theta ) \; = \; - \theta \; \;   \forall r \in ( 0 , + \infty ) , \forall \theta \in [ 0 , \frac{\pi}{2}
    ]
\end{equation}

 we have that:
\begin{equation}
    L_{1} ( r , \theta , \dot{r} , \dot{\theta} ) \; = \;
    \frac{m}{2} ( \dot{r}^{2} + r^{2} \dot{\theta}^{2} ) + \theta
    \; \; r \in ( 0 , + \infty ) , \theta \in [ 0 , \frac{\pi}{2}
    ] , \dot{r} \in \mathbb{R} , \dot{\theta} \in \mathbb{R}
\end{equation}
\begin{equation}
    H_{1} ( r , \theta , p_{r} , p_{\theta} ) \; = \;
    \frac{p_{r}^{2}}{2m} + \frac{p_{\theta}^{2} }{2 m r^{2}} - \theta  \; \; r \in ( 0 , + \infty ) , \theta \in [ 0 ,
    \frac{\pi}{2} ] , p_{r} \in \mathbb{R} , p_{\theta} \in \mathbb{R}
\end{equation}
where:
\begin{equation}
     H_{1} ( r , \theta , p_{r} , p_{\theta} ) \; = \; p_{r}
     \dot{r} + p_{\theta} \dot{\theta} -  L_{1} ( r , \theta , \dot{r} , \dot{\theta} )
\end{equation}
\begin{equation}
    p_{r} \; = \; \frac{\partial L_{1}}{ \partial \dot{r} } \; =
    \; m \dot{r}
\end{equation}
\begin{equation}
    p_{\theta} \; = \; \frac{\partial L_{1}}{ \partial \dot{\theta}} \; =
    \; m r^{2} \dot{\theta}
\end{equation}
Hamilton's equations are:
\begin{equation}
     \dot{p_{r}} \; = \; - \frac{\partial H_{1}}{ \partial r } \; = \; \frac{ p_{\theta}^{2} }{m r^{3}}
\end{equation}
\begin{equation}
    \dot{p_{\theta}} \; = \; - \frac{\partial H_{1}}{ \partial \theta } \; = \; 1
\end{equation}
\begin{equation}
    \dot{r} \; = \;  \frac{\partial H_{1}}{ \partial p_{r} } \; = \; \frac{p_{r}}{m}
\end{equation}
\begin{equation}
     \dot{\theta} \; = \; \frac{\partial H_{1}}{ \partial p_{\theta} } \; = \; \frac{p_{\theta}}{m r^{2}}
\end{equation}
and hence:
\begin{equation}
    p_{\theta} (t) \; = \; p_{\theta}(0) + t
\end{equation}
\begin{equation}
   \dot{p_{r}} (t) \; = \; \frac{ (p_{\theta} (0) + t)^{2}  }{m r^{3}}
\end{equation}

\bigskip

Given a set $ \{ V_{i} \}_{i=1}^{n} $ of local energy potentials
for f let us observe that if  $ U_{i} \cap U_{j} \; \neq \;
\emptyset $ then the map $ t_{ij} : U_{i} \cap U_{j} \mapsto
\mathbb{R} $:
\begin{equation}
    t_{ij}(q) \; := \; \exp ( c_{ij}(q) )
\end{equation}
is constant.

Furthermore equation \ref{eq:first condition} , equation
\ref{eq:second condition} and equation \ref{eq:third condition}
imply that:
\begin{enumerate}
    \item
\begin{equation} \label{eq:first condition exponentiated}
    t_{ii} (q) \; = \; 1
\end{equation}
    \item
\begin{equation}  \label{eq:second condition exponentiated}
    t_{ij} (q) \; = \; t_{ji}(q)^{- 1}
\end{equation}
    \item if  $ U_{i} \cap U_{j}  \cap U_{k} \; \neq \;
\emptyset $ then:
\begin{equation} \label{eq:third condition exponentiated}
    t_{ij} (q) \cdot t_{jk}(q) \; = \; t_{ik}(q)
\end{equation}
\end{enumerate}

Let us now observe that equation \ref{eq:first condition
exponentiated} , \ref{eq:second condition exponentiated} and
\ref{eq:third condition exponentiated} are nothing but the
consistency conditions satisfied by the transition functions of a
fibre bundle.

We can consequentially follow the  strategy indicated in the
section 9.2.2 of \cite{Nakahara-03} to construct a principal
bundle $P(M , \mathbb{R} ) $ given M , $ \{ U_{i} \}_{i=1}^{n} $,
the transition functions $ \{ t_{ij} (q) \} $ (that in our case
are constant) and the structure group $ \mathbb{R} $:

let us start introducing the set:
\begin{equation}
    X \; := \; \cup_{i=1}^{n} U_{i} \, \times \, \mathbb{R}
\end{equation}
Given $ ( q_{1} , a_{1} ) \in U_{i} \times \, \mathbb{R} $ and   $
( q_{2} , a_{2} ) \in U_{j} \times \, \mathbb{R} $ and introduced
the following equivalence relation:
\begin{equation}
    ( q_{1} , a_{1} ) \sim ( q_{2} , a_{2} ) \; := \; q_{1} =
    q_{2} \, \wedge \, a_{2} = t_{ij} (q) a_{1}
\end{equation}
the total space P is simply defined as the quotient set:
\begin{equation}
 P \; := \; \frac{X}{\sim}
\end{equation}
Denoted an element of P as $ [ (q , a) ] $ the projection $ \pi :
P \mapsto M $ is defined as:
\begin{equation}
    \pi (  [ (q , a) ] ) \; := \; q
\end{equation}
The local trivialization $ \phi_{i} : U_{i} \times \mathbb{R}
\mapsto \pi^{-1}(U_{i}) $ is defined as:
\begin{equation}
  \phi_{i} : ( q , a) \mapsto [ ( q , a) ]
\end{equation}

\smallskip

A set of local energy potentials of the locally conservative
force's field f may be interpreted as a set of local sections of $
P(M , \mathbb{R})$ by considering a set of maps $ \{ \tilde{V}_{i}
\}_{i=1}^{n} $ such that:
\begin{enumerate}
    \item
\begin{equation}
   \tilde{V}_{i} \in \Gamma (U_{i} , P) \; \; \forall i \in \{ 1 ,
   \cdots , n \}
\end{equation}
    \item
\begin{equation}
   \tilde{V}_{i} (q) \; := \; \phi_{i} ( V_{i}(q) , 0 ) \; = \; [ ( V_{i}(q) , 0
   ) ]
\end{equation}
\end{enumerate}

\smallskip

From the principal bundle $ P( M , \mathbb{R} ) $ one can then
naturally derive:
\begin{enumerate}
    \item the principal bundle $ P_{L} ( TM , \mathbb{R} )
$ defined as:
\begin{equation}
    P_{L} \; := \; \frac{ \cup_{i=1}^{n} T U_{i} \, \times \, \mathbb{R} }{\sim_{L}}
\end{equation}
where:
\begin{multline}
    ( ( q_{1} , v_{1} ) , a_{1} ) \sim_{L} ( ( q_{2} , v_{2} ) ,
    a_{2}) \; := \\
     ( q_{1} , a_{1} ) \sim ( q_{2} , a_{2} ) \; \;
    \forall v_{1} \in T_{q_{1}} U_{i} , \forall v_{2} \in T_{q_{2}}
    U_{j} , \forall ( q_{1} , a_{1} ) \in U_{i} \times \,
    \mathbb{R} , \forall ( q_{2} , a_{2} ) \in U_{j} \times \, \mathbb{R}
\end{multline}
    \item the principal bundle $ P_{H} ( T^{\star} M , \mathbb{R} )
$ defined as:
\begin{equation}
    P_{H} \; := \; \frac{ \cup_{i=1}^{n} T^{\star} U_{i} \, \times \, \mathbb{R} }{\sim_{H}}
\end{equation}
where:
\begin{multline}
    ( ( q_{1} , p_{1} ) , a_{1} ) \sim_{H} ( ( q_{2} , p_{2} ) ,
    a_{2}) \; := \\
     ( q_{1} , a_{1} ) \sim ( q_{2} , a_{2} ) \; \;
    \forall p_{1} \in T^{\star}_{q_{1}} U_{i} , \forall p_{2} \in T^{\star}_{q_{2}}
    U_{j} , \forall ( q_{1} , a_{1} ) \in U_{i} \times \,
    \mathbb{R} , \forall ( q_{2} , a_{2} ) \in U_{j} \times \, \mathbb{R}
\end{multline}
\end{enumerate}

The set of local lagrangians $ \{ L_{i} \}_{i=1}^{n} $ for the
locally conservative physical system $ PS_{f} $ may be interpreted
as a set of local sections of the bundle $ P_{L} ( TM , \mathbb{R}
) $ by considering the set of maps $ \{ \tilde{L}_{i} \}_{i=1}^{n}
$ such that:
\begin{enumerate}
    \item
\begin{equation}
   \tilde{L}_{i} \in \Gamma (T U_{i} , P_{L}) \; \; \forall i \in \{ 1 ,
   \cdots , n \}
\end{equation}
    \item
\begin{equation}
   \tilde{L}_{i} (q, \dot{q}) \; := \; \frac{m}{2} |\dot{q} |_{g}^{2} -
   \tilde{V}_{i}(q)
\end{equation}
\end{enumerate}

In an analogous way the set of local hamiltonians $ \{ H_{i}
\}_{i=1}^{n} $ for the locally conservative physical system $
PS_{f} $ may be interpreted as a set of local sections of the
bundle $ P_{H} ( TM , \mathbb{R} ) $ by considering the set of
maps $ \{ \tilde{H}_{i} \}_{i=1}^{n} $ such that:

\begin{enumerate}
    \item
\begin{equation}
   \tilde{H}_{i} \in \Gamma (T^{\star} U_{i} , P_{H}) \; \; \forall i \in \{ 1 ,
   \cdots , n \}
\end{equation}
    \item
\begin{equation}
   \tilde{H}_{i} (p,q) \; := \; \frac{|p |_{g}^{2} }{2 m}  +
   \tilde{V}_{i}(q)
\end{equation}
\end{enumerate}

\smallskip

\begin{remark}
\end{remark}
If the locally conservative force's field f is conservative then
the principal bundles $ P ( M , \mathbb{R} ) $ , $ P_{L} ( TM ,
\mathbb{R} ) $ and $ P_{H} ( T^{\star} M , \mathbb{R} ) $ are
trivial and hence they admit global sections that are,
respectively, a globally defined energy potential, a globally
defined lagrangian and a globally defined hamiltonian.

If, contrary, the locally conservative force's field f is not
conservative then the principal bundles $ P ( M , \mathbb{R} ) $ ,
$ P_{L} ( TM , \mathbb{R} ) $ and $ P_{H} ( T^{\star} M ,
\mathbb{R} ) $ are not trivial and hence they don't admit global
sections.

Anyway the same existence of these bundles is sufficient to
guarantee that all the local descriptions of $ PS_{f} $ define in
a consistent way a global dynamics.

\bigskip

\begin{remark}
\end{remark}
Let us remark that in a locally conservative but not conservative
physical system the mechanical energy is conserved locally but not
globally.

As a whole such a system is dissipative, converting a portion of
work into heat \footnote{Of course, as we have already remarked,
the Conservation of Mechanical Energy, lost in terms of the
time-invariance of  a globally defined hamiltonian, is replaced by
the more general Conservation of Energy (in all its forms)
guaranteed by the First Principle of Thermodynamics.}.

The physical source of such a dissipation is, in the final
analysis, the topological non-triviality of the involved manifold.

\bigskip

\begin{example} \label{ex:prototype5}
\end{example}
In the framework of the example \ref{ex:prototype1}, of the
example \ref{ex:prototype2}, of the example \ref{ex:prototype3}
and of the example \ref{ex:prototype4} let us construct the
principal bundle $ P ( \mathbb{R}^{2} - \{ \vec{0} \} , \mathbb{R}
) $.

Let us observe first of all that:
\begin{enumerate}
    \item
\begin{equation}
    U_{1} \cap U_{2} \; = \; \{ ( 0 ,y) \; y \in (0, + \infty ) \}
\end{equation}
\begin{equation}
    t_{12}(q) \; = \; \exp( + \pi ) \; \; \forall q \in  U_{1} \cap U_{2}
\end{equation}
    \item
\begin{equation}
    U_{2} \cap U_{3} \; = \; \{ ( x ,0) \; x \in (- \infty , 0 ) \}
\end{equation}
\begin{equation}
    t_{23}(q) \; = \; \exp( - \pi ) \; \; \forall q \in  U_{2} \cap U_{3}
\end{equation}
    \item
\begin{equation}
    U_{3} \cap U_{4} \; = \; \{ ( 0 ,y) \; y \in ( - \infty , 0 ) \}
\end{equation}
\begin{equation}
    t_{34}(q) \; = \; \exp( + \pi ) \; \; \forall q \in  U_{3} \cap U_{4}
\end{equation}
    \item
\begin{equation}
    U_{4} \cap U_{1} \; = \; \{ ( x ,0) \; x \in (0, + \infty ) \}
\end{equation}
\begin{equation}
    t_{41}(q) \; = \; \exp( - \pi ) \; \; \forall q \in  U_{4} \cap U_{1}
\end{equation}
\end{enumerate}

Introduced the set:
\begin{equation}
    X \; := \; \cup_{i=1}^{4} U_{i} \times \mathbb{R}
\end{equation}
the equivalence relation involved in the definition of $ P :=
\frac{X}{\sim} $ is the following:
\begin{enumerate}
    \item given $ ( q_{1} , a_{1} ) \in U_{1} \times \mathbb{R} $
    and $ ( q_{2} , a_{2} ) \in U_{2} \times \mathbb{R} $:
\begin{equation}
   ( q_{1} , a_{1} ) \sim  ( q_{2} , a_{2} ) \; \Leftrightarrow \;
   q_{1} = q_{2} \; \wedge \; a_{2} = \exp ( + \pi ) a_{1}
\end{equation}
    \item given $ ( q_{1} , a_{1} ) \in U_{2} \times \mathbb{R} $
    and $ ( q_{2} , a_{2} ) \in U_{3} \times \mathbb{R} $:
\begin{equation}
   ( q_{1} , a_{1} ) \sim  ( q_{2} , a_{2} ) \; \Leftrightarrow \;
   q_{1} = q_{2} \; \wedge \; a_{2} = \exp ( - \pi ) a_{1}
\end{equation}
    \item  given $ ( q_{1} , a_{1} ) \in U_{3} \times \mathbb{R} $
    and $ ( q_{2} , a_{2} ) \in U_{4} \times \mathbb{R} $:
\begin{equation}
   ( q_{1} , a_{1} ) \sim  ( q_{2} , a_{2} ) \; \Leftrightarrow \;
   q_{1} = q_{2} \; \wedge \; a_{2} = \exp ( + \pi ) a_{1}
\end{equation}
    \item given $ ( q_{1} , a_{1} ) \in U_{4} \times \mathbb{R} $
    and $ ( q_{2} , a_{2} ) \in U_{1} \times \mathbb{R} $:
\begin{equation}
   ( q_{1} , a_{1} ) \sim  ( q_{2} , a_{2} ) \; \Leftrightarrow \;
   q_{1} = q_{2} \; \wedge \; a_{2} = \exp ( - \pi ) a_{1}
\end{equation}
\end{enumerate}

The set of local sections $ \{ \tilde{V}_{i} \in \Gamma (U_{i},P)
\}_{i=1}^{4} $ is:
\begin{enumerate}
    \item
\begin{equation}
    \tilde{V}_{1} ( x ,y ) \; = \;  [ ( - \arctan ( \frac{y}{x}),0)
    ] \; \; \forall ( x,y) \in U_{1} : ( x,y)  \notin U_{1} \cap U_{2}
\end{equation}
\begin{equation}
   \tilde{V}_{1} ( x ,y ) \; = \;  [ ( + \frac{\pi}{2} , 0 )]  \; \; \forall ( x,y)  \in U_{1} \cap U_{2}
\end{equation}
    \item
\begin{equation}
    \tilde{V}_{1} ( x ,y ) \; = \; [ ( - \arctan \frac{y}{x} , 0)
    ] \; \; \forall ( x,y) \in U_{2} : ( x,y)  \notin U_{2} \cap U_{3}
\end{equation}
\begin{equation}
      \tilde{V}_{1} ( x ,y ) \; = \;  [ ( - \frac{\pi}{2}, 0 )] \; \; \forall ( x,y)  \in U_{2} \cap U_{3}
\end{equation}
    \item
\begin{equation}
     \tilde{V}_{3} ( x ,y ) \; = \;  [ ( - \arctan \frac{y}{x},0)
     ] \; \; \forall ( x,y) \in U_{3} : ( x,y)  \notin U_{3} \cap U_{4}
\end{equation}
\begin{equation}
     \tilde{V}_{3} ( x ,y ) \; = \; [ ( + \frac{\pi}{2}, 0 )]  \;
     \;  \forall ( x,y)  \in U_{3} \cap U_{4}
\end{equation}
    \item
\begin{equation}
    \tilde{V}_{4} ( x ,y )  \; = \;  [ ( - \arctan \frac{y}{x} , 0)
    ]  \; \; \forall ( x,y) \in U_{4} : ( x,y)  \notin U_{4} \cap U_{1}
\end{equation}
\begin{equation}
       \tilde{V}_{4} ( x ,y )  \; = \;  [ ( - \frac{\pi}{2}, 0 )] \;
     \;  \forall ( x,y)  \in U_{4} \cap U_{1}
\end{equation}
\end{enumerate}
from which the construction of $ \{ \tilde{L}_{i} \in \Gamma ( T
U_{i} , P_{L} ) \}_{i=1}^{4} $ and of $ \{ \tilde{H}_{i} \in
\Gamma ( T^{\star} U_{i} , P_{H} ) \}_{i=1}^{4} $ may be
immediately derived.

\bigskip

Let us now introduce an alternative approach to the formulation of
the Analytic Mechanics of $ PS_{f} $ essentially consisting in the
lifting to the universal covering space $ \tilde{M} $.

Given a set $ \{ V_{i} \}_{i=1}^{n} $ of local energy potentials
for the locally conservative force's field f let us introduce the
following:
\begin{definition}
\end{definition}
\emph{lift of $ \{ V_{i} \}_{i=1}^{n} $ to the universal covering
space:}

the set of functions $ \{ \tilde{V}_{i} \}_{i=1}^{n} $ such that:
\begin{enumerate}
    \item $  \tilde{V}_{i} \in \Gamma ( U_{i} , \tilde{M} )
    $ where $ \Gamma ( U_{i} , \tilde{M} ) $ is the set of the
    local sections defined on $ U_{i} $ of the principal bundle $ \tilde{M} ( M , \pi_{1}(M) ) $
    \item
\begin{equation}
    \tilde{V}_{i} ( \tilde{q} ) \; := \; V_{i}( \pi (  \tilde{q}))
    \; \; \forall i \in \{ 1 , \cdots , n \}
\end{equation}
\end{enumerate}

Given the set $ \{ L_{i} \}_{i=1}^{n} $ of local lagrangians
associated to the set $  \{ V_{i} \}_{i=1}^{n} $ of local energy
potentials let us introduce the following:

\begin{definition}
\end{definition}
\emph{lift of $ \{ L_{i} \}_{i=1}^{n} $ to the universal covering
space:}

the set of functions  $ \{ \check{L}_{i} \}_{i=1}^{n} $ such that:
\begin{enumerate}
    \item  $ \check{L}_{i} \in \Gamma ( U_{i} , T \tilde{M} ) $ where $ \Gamma ( U_{i} , \tilde{M} ) $ is the set of the
    local sections defined on $ U_{i} $ of the tangent bundle of $
    \tilde{M} $
    \item
\begin{equation}
  \check{L}_{i} ( \tilde{q} , \dot{\tilde{q}}) \; := \; \frac{m}{2}
   | \dot{\tilde{q}} |_{g}^{2} - \tilde{V}_{i} ( \tilde{q} )
\end{equation}
where $ \tilde{g} $ is the lift of the riemannian metric g to $
\tilde{M} $.
\end{enumerate}

In an analogous way, given the set  $ \{ H_{i} \}_{i=1}^{n} $ of
local hamiltonians associated to the set $  \{ V_{i} \}_{i=1}^{n}
$ of local energy potentials  $ \{ V_{i} \}_{i=1}^{n} $:
\begin{definition}
\end{definition}
\emph{lift of $ \{ H_{i} \}_{i=1}^{n} $ to the universal covering
space:}

the set of functions  $ \{ \check{H}_{i} \}_{i=1}^{n} $ such that:
\begin{enumerate}
    \item  $ \check{H}_{i} \in \Gamma ( U_{i} , T^{\star} \tilde{M} ) $ where $ \Gamma ( U_{i} , \tilde{M} ) $ is the set of the
    local sections defined on $ U_{i} $ of the cotangent bundle of $
    \tilde{M} $
    \item
\begin{equation}
  \check{H}_{i} ( \tilde{q} , \tilde{p}) \; := \;
\frac{ | \tilde{p} |_{g}^{2}}{2 m}  + \tilde{V}_{i} ( \tilde{q} )
\end{equation}
\end{enumerate}

\bigskip

\begin{example} \label{ex:prototype6}
\end{example}
In the framework of the example \ref{ex:prototype1},  of the
example \ref{ex:prototype1}, of the example \ref{ex:prototype3} ,
of the example \ref{ex:prototype4} and  of the example
\ref{ex:prototype5} let us endow $ \mathbb{R}^{2} $ with  the
following binary inner operations:
\begin{equation}
     ( x_{1} , y_{1} ) + ( x_{2} , y_{1} ) \; := \;  ( x_{1} + x_{2} , y_{1} + y_{2} )
\end{equation}
\begin{equation}
    ( x_{1} , y_{1} ) \cdot ( x_{2} , y_{1} ) \; := \; ( x_{1}
    x_{2} - y_{1} y_{2} , x_{1} y_{2} + x_{2} y_{1} )
\end{equation}
and let us make the usual identification $ ( \mathbb{R}^{2} , + ,
\cdot ) \; \equiv \; \mathbb{C}$.

Let us then observe that:
\begin{equation}
    f^{\sharp} ( z) \; = \; \frac{i}{z}
\end{equation}
(analytic on $ \mathbb{C} - \{ 0 \} $ with a pole in $z=0$) and
that given a loop $ c : [ 0, 1] \mapsto M $ and remembering that $
Res( \frac{1}{z} , 0 ) \, = \, 1 $ one has that:
\begin{equation}
    \oint_{c} f(z) dz \; = \; - 2 \pi n_{winding} ( c, \vec{0} ) \; = \;  - W ( f, c)
\end{equation}

Since $ M = \mathbb{C}^{\times} :=  \mathbb{C} - \{ 0 \} $ is the
punctured complex plane it follows that (see the example 2 of the
section 2.9 of \cite{Shokurov-94}):
\begin{equation}
    \tilde{M} \; = \; \mathbb{C}
\end{equation}
the universal covering map $ \pi :  \mathbb{C} \mapsto
\mathbb{C}^{\times} $ being the exponential:
\begin{equation}
   \pi (z) = \exp( z)
\end{equation}

Obviously:
\begin{equation}
    \pi_{1} (  \mathbb{C}^{\times} ) \; = \; \mathbb{Z}
\end{equation}

Considered the following action of $  \pi_{1} (
\mathbb{C}^{\times} ) $ on $ \mathbb{C} $:
\begin{equation}
    n : z \mapsto z + 2 \pi n i
\end{equation}
one has that:
\begin{equation}
  \mathbb{C}^{\times} \; = \; \frac{\mathbb{C}}{\mathbb{Z}}
\end{equation}

\bigskip

\begin{remark}
\end{remark}
Let us recall that, in general, given a discrete group G acting on
a manifold M the quotient space $ \frac{M}{G} $ is not a manifold,
but it is only an orbifold (according to the definition given by
William Thurston about which we demand to the appendix E of
\cite{Milnor-06}).

If, as in our case, such an action is free then the quotient space
is also a manifold.

\bigskip

\begin{figure}
  \includegraphics[scale=.5]{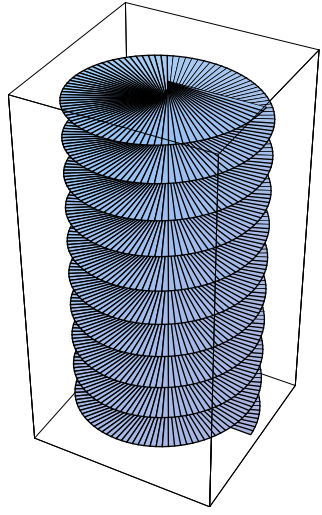}\\
  \caption{The intersection of $ \Sigma_{\log} $ and a cylinder centered in the origin $ \vec{0}$ of $ \mathbb{R}^{3}$.} \label{fig:helix}
\end{figure}

 Instead of introducing a lift of the local energy potentials $ \{ V_{i} \}_{i=1}^{4}$
 to the universal covering space $ \tilde{M} = \mathbb{C} $ let us
 observe that the angle Arg(z) may be seen as the complete
 analytic function $ \theta : \Sigma_{\log} \mapsto \mathbb{C} $:
 \begin{equation}
    \theta ( z ) \; := \frac{1}{i}  \log ( \frac{z}{ | z|})
\end{equation}

where $ \Sigma_{\log} $ is  the  Riemann surface of the logarithm
(see the section \ref{sec:The double meaning of the locution
"Riemann Surface"} and the figure \ref{fig:helix}) so that it is
natural to introduce the \emph{Riemann-surface lifting} $
\tilde{V} : \Sigma_{\log} \mapsto \mathbb{C} $:
\begin{equation}
 \breve{V}  ( z ) \; := - \theta (z)
\end{equation}

The dynamics of $ PS_{f} $ should then be describable trough the
lagrangian $ \breve{L}  : T \Sigma_{\log} \mapsto \mathbb{R} $:
\begin{equation}
   \breve{L} ( z , \dot{z} ) \; := \; \frac{1}{2} | \dot{z}
   |_{\breve{\delta}}^{2} -  \breve{V}(z)
\end{equation}
where $ \breve{\delta} $ is the riemannian metric on $
\Sigma_{log} $ induced by the euclidean metric on the plane.
\newpage
\appendix
\section{Passing from the language of vector calculus to the language of differential
forms} \label{sec:Passing from the language of vector calculus to
the language of differential forms}

Let us consider the three-dimensional euclidean manifold $
\mathbb{E}^{3} := ( \mathbb{R}^{3} , \delta := dx \otimes dx + d y
\otimes d y +  dz \otimes dz     ) $.

Introduced the canonical basis $ \{ \vec{e}_{1} , \vec{e}_{2} ,
\vec{e}_{3} \} $ of $ \mathbb{R}^{3} $ defined as:
\begin{equation}
    ( \vec{e}_{i} )_{j} \; := \; \delta_{i,j} \; \; i,j= 1 ,
    \cdots , 3
\end{equation}
let us adopt the following Sharp and Flat notation of
\cite{Marsden-Ratiu-99}:

Given the 1-form $ \alpha = \alpha_{x} dx + \alpha_{y} dy +
\alpha_{z} dz \in \Omega^{1} ( \mathbb{R}^{3} ) $ let us introduce
the following:
\begin{definition}
\end{definition}
\emph{vector field associated to $ \alpha $:}
\begin{equation}
    \alpha^{\sharp} \; := \; \alpha_{x}  \vec{e}_{1} +  \alpha_{y}
    \vec{e}_{2} + \alpha_{z}
    \vec{e}_{3}
\end{equation}
Given, contrary, the vector field $ \vec{v} := v_{x} \vec{e}_{1} +
v_{y} \vec{e}_{2}+ v_{z} \vec{e}_{3} $:
\begin{definition}
\end{definition}
\emph{1-form associated to $ \vec{v} $:}
\begin{equation}
    \vec{v}^{\flat} \; := \; v_{x} dx +
v_{y} dy + v_{z} dz
\end{equation}

Then:
\begin{proposition}
\end{proposition}
\begin{enumerate}
    \item Cross Product:
\begin{equation}
 \vec{v} \wedge \vec{w} \; = \; [ \star ( \vec{v}^{\flat} \wedge
 \vec{w}^{\flat}) ]^{\sharp}
\end{equation}
  \item Scalar Product:
\begin{equation}
    ( \vec{v} \cdot \vec{w} ) dx \wedge dy \wedge dz \; = \;
    \vec{v}^{\flat} \wedge \star ( \vec{w}^{\flat} )
\end{equation}
    \item Gradient:
\begin{equation}
   \vec{ \nabla} f \; := \; grad f \; = \; ( d f)^{\sharp}
\end{equation}
    \item Curl:
\begin{equation}
   \vec{ \nabla} \wedge \vec{v} \; := \; curl \vec{v} \; = \; [
   \star ( d \vec{v}^{\flat} )]^{\sharp}
\end{equation}
    \item Divergence:
\begin{equation}
  \vec{ \nabla} \cdot \vec{v} \; := \; div \vec{v} \; = \; \star d
  ( \star \vec{v}^{\flat} )
\end{equation}
\end{enumerate}
where $ \star $ is the Hodge Star operator of $ \mathbb{E}^{3} $
satisfying the following equations:
\begin{equation}
    \star 1 \; = \; dx \wedge dy \wedge dz
\end{equation}
\begin{equation}
    \star dx \; = \; dy \wedge dz
\end{equation}
\begin{equation}
    \star dy \; = \; - d x \wedge dz
\end{equation}
\begin{equation}
    \star d z \; = \; dx \wedge dy
\end{equation}
\begin{equation}
    \star ( dx \wedge dy ) \; = \; dz
\end{equation}
\begin{equation}
    \star ( dx \wedge dz ) \; = \; - dy
\end{equation}
\begin{equation}
    \star ( dy \wedge d z ) \; = \; d x
\end{equation}
\begin{equation}
    \star ( dx \wedge dy \wedge dz ) \; = \; 1
\end{equation}
\newpage
\section{The double meaning of the locution "Riemann Surface"} \label{sec:The double meaning of the locution "Riemann Surface"}
The locution "Riemann surface" appears in the Theoretical Physics'
literature in two distinct meanings:
\begin{enumerate}
    \item as a bidimensional submanifold of $ \mathbb{R}^{3} $
    satisfying suitable conditions
    \item as the domain of definition (a finite or countable set of copies of $ \mathbb{C} $ patched together through a suitable number of "cut and paste" operations) on which a multivalued
    complex function of a complex variable becomes single-valued (we will refer to this approach as to the carpenter's
    definition of a Riemann surface)
\end{enumerate}
The deep link existing between the two meanings is something
usually taken for granted though it is never clarified.

Actually its full comprehension requires to give up the
carpenter's definition  and to introduce more advanced
mathematical concepts that we think it may be appropriate to
review here (demanding to \cite{Conway-78}, \cite{Conway-95},
\cite{Farkas-Kra-92} for further details).

\bigskip

\begin{example}
\end{example}
Let us consider the carpenter's definition of the Riemann surface
of the logarithm.

Given  $ z_{1} , z_{2} \in \mathbb{C} $:

\begin{definition}
\end{definition}
\emph{$ z_{1} $ is a logarithm of $ z_{2}$:}
\begin{equation}
    \exp ( z_{1} ) \; = \; z_{2}
\end{equation}

Given $ r \in ( 0 , + \infty) $ and $ \theta \in [ 0 ,2 \pi ) $
one may easily verify that the set of the logarithms of $ r \exp (
i \theta ) $ is $ \{ \log(r) + i ( \theta + 2 \pi n ) \; n \in
\mathbb{Z} \} $.

The carpenter's definition of $ \Sigma_{log} $ proceeds in the
following way:
\begin{enumerate}
    \item one considers a sequence $ \{  sheet(n) , n \in
    \mathbb{Z} \} $ such that:
\begin{equation}
    sheet(n) \; := \; \mathbb{C} - \{ 0 \} \; \; \forall n \in \mathbb{Z}
\end{equation}
   \item for every $ n \in \mathbb{Z} $ one defines the map $ \phi_{n} : sheet(n) \mapsto
   \mathbb{C} $:
\begin{equation}
  \phi_{n} ( z ) \; := \; \log ( | z | ) + i ( Arg(z) + 2 \pi n )
\end{equation}
  \item one defines $ \log : \cup_{n \in \mathbb{Z}} sheet(n)
  \mapsto \mathbb{C} $ such that:
\begin{equation}
  \log|_{sheet(n)} \; := \;  \phi_{n}
\end{equation}
    \item one "cuts" each sheet(n) along the negative real
    semiaxis $ ( - \infty , 0) $ so that such an interval is
    replaced with two copies of it that one calls the upper border
    and the lower border of the cut sheet, i.e.:
\begin{equation}
    border ( n , + ) \; := \; ( - \infty , 0 )
\end{equation}
\begin{equation}
    border ( n , - ) \; := \; ( - \infty , 0 )
\end{equation}
\begin{equation}
    sheet(n) = \mathbb{C} - \{ 0 \} \; \rightarrow \; sheet(n) := ( \mathbb{C}
    - ( - \infty , 0 ] \cup   border ( n , + ) \cup border ( n , -
    ) \; \; \forall n \in \mathbb{Z}
\end{equation}
    \item one "welds together" $ border( n , +) $ and $ border(
    n+1 , - )$ by making the identifications:
\begin{equation}
border( n , +) \; \equiv \; border ( n+1 , -) \; \; \forall n \in
\mathbb{Z}
\end{equation}
obtaining a surface $ \Sigma_{log} $ on which the logarithm
results defined.
\end{enumerate}

\bigskip

Given a topological space $ ( M , \mathcal{T} ) $ \cite{Nash-94}:
\begin{definition} \label{def:sheaf of germs}
\end{definition}
\emph{sheaf of germs $ \mathcal{S} $  on $ ( M , \mathcal{T} ) $:}

The assignment to each $ U \in \mathcal{T} $ of a group $
\mathcal{S} (U)$  (called the\emph{ group of sections of $
\mathcal{S} $ over U}) such that:
\begin{enumerate}
    \item Given $ U , V  \in \mathcal{T} $ such that $
    U \subseteq V $ there exist a map $ r_{U}^{V}
    : \mathcal{S} ( V) \mapsto  \mathcal{S} (U) $ (called
    the \emph{restriction map from $ V $ to $ U $} such
    that:
\begin{equation}
  r_{U}^{U} \; = \; \mathbb{I} \; \; \forall U \in  \mathcal{T}
\end{equation}
\begin{equation}
   ( U \subseteq V \subseteq W \; \Rightarrow \;
    r_{U}^{W} \, = \, r_{U}^{V} \circ
    r_{V}^{W} ) \; \; \forall U, V, W \in \mathcal{T}
\end{equation}
    \item Given $ U \in \mathcal{T} $ such that $ U =
    \cup_{i} U_{i} \; :  \; U_{i} \in \mathcal{T} \; \forall
    i $:
\begin{equation}
    s_{1} , s_{2} \in  \mathcal{F} (U) \, : \, r_{U_{i}}^{U} (
    s_{1}) = r_{U_{i}}^{U} (
    s_{2}) \, \forall i \; \Rightarrow \; s_{1} = s_{2}
\end{equation}
\begin{equation}
    r_{U_{i}\cap U_{j}}^{U_{i}} ( s_{i} ) \, = \,  r_{U_{i}\cap U_{j}}^{U_{j}} ( s_{j}
    ) \, \forall i, j \; \Rightarrow \; \exists ! s \in
    \mathcal{F} (U) \, : \, r_{U_{i}}^{U}(s) = s_{i} \, \forall i
\end{equation}
\end{enumerate}
Given a sheaf $ \mathcal{S} $ on $ ( M , \mathcal{T} ) $ and given
$ x \in M $:
\begin{definition} \label{def:set of the germs of a sheaf in a point of the base space}
\end{definition}
\emph{set of the germs of $  \mathcal{S} $ at x:}
\begin{equation}
    \mathcal{S}_{x} \; := \; \lim_{ \rightarrow x \in U} \mathcal{F} (U)
\end{equation}
where $  \lim_{\rightarrow x \in U}  $ is a \emph{direct limit}.

Suitable generalizations of the definition \ref{def:sheaf of
germs} have been the starting point from which Alexander
Grothendieck's extraordinary abstraction's skills have led to
strongly remarkable results for a taste of which we demand to
\cite{Mc-Lane-Moerdijk-92}.

Though fibre-bundles are nowadays considered as part of the
differential geometric tool-bag of a theoretical physicist the
same cannot be said about sheaves of germs (as  remarked by Chris
Isham in the section 5.1.4 of \cite{Isham-99} and still valid
after almost a decade).

One of the reasons is that every sheaf of germs is the sheaf of
germs of the local sections of a suitable fibre bundle.

The notion of a \emph{sheaf of germs} plays, anyway, a crucial
role in the rigorous definition of a Riemann surface of a complex
function of a complex variable.

\begin{definition}
\end{definition}
\emph{region of $ \mathbb{C}$}:
\begin{equation}
    U \subset \mathbb{C} \text{ open and connected}
\end{equation}

\begin{definition}
\end{definition}
\emph{function element:}

a couple $ ( f , G ) $ such that:
\begin{enumerate}
    \item G is a region of $ \mathbb{C}$
    \item f is an analytic function on G
\end{enumerate}

Given a function element $ ( f , G ) $ and a point $ a \in G$:

\begin{definition}
\end{definition}
\emph{germ of f at a:}
\begin{equation}
    [ f ]_{a} \; := \; \{ ( g , D ) \text{ function element } : a
    \in D \; \wedge \; \exists \,  U \text{ neighborhood of a } : g(z)
    = f(z) \, \forall z \in U \}
\end{equation}

Given a path $ \gamma : [ 0, 1] \mapsto \mathbb{C} $ and a family
$ \{ ( f_{t} , D_{t} ) , t \in [ 0,1] \} $ of function elements:
\begin{definition}
\end{definition}
\emph{$ ( f_{1} , D_{1})$ is the analytic continuation of $ (
f_{0} , D_{0})$ along $ \gamma $:}
\begin{equation}
    \forall t \in [0,1] \; \exists \delta > 0 \; : \; ( | s - t| <
    \delta \, \Rightarrow \, \gamma(s) \in D_{t} ) \; \wedge \; [
    f_{s} ]_{\gamma(s)} = [
    f_{t} ]_{\gamma(s)}
\end{equation}

\begin{proposition} \label{prop:property behind the analytic continuation of germs}
\end{proposition}

\begin{hypothesis}
\end{hypothesis}
\begin{center}
$ \gamma : [ 0 , 1 ] \mapsto \mathbb{C} \; : \; \gamma(0) = a
\wedge \gamma(1) = b $ path
\end{center}
\begin{center}
  $ \{ ( f_{t} , D_{t}) \, t \in [ 0 ,1 ] \} \; , \;  \{ ( g_{t} , B_{t}) \, t \in [ 0 ,1 ]
  \} $ analytic continuations of $ ( f_{0} , D_{0}) $ along $
  \gamma$ such that $ [ f_{0} ]_{a} \, = \; [ g_{0} ]_{a} $
\end{center}
\begin{thesis}
\end{thesis}
\begin{equation*}
    [ f_{1} ]_{b} \; = \; [ g_{1} ]_{b}
\end{equation*}

Given a path  $ \gamma : [ 0 , 1 ] \mapsto \mathbb{C} \; : \;
\gamma(0) = a \wedge \gamma(1) = b $ and given $ \{ ( f_{t} ,
D_{t}) \, t \in [ 0 ,1 ] \} \; , \;  \{ ( g_{t} , B_{t}) \, t \in
[ 0 ,1 ] \} $ analytic continuations of $ ( f_{0} , D_{0}) $ along
$ \gamma $ such that $ [ f_{0} ]_{a} \, = \; [ g_{0} ]_{a} $
Proposition \ref{prop:property behind the analytic continuation of
germs} justifies the following:
\begin{definition}
\end{definition}
\emph{analytic continuation of $ [ f_{0} ]_{a} $ along $ \gamma
$:}
\begin{center}
 $[ f_{1}]_{b}$
\end{center}

\smallskip

Given a function element $ ( f , G ) $:
\begin{definition}
\end{definition}
\emph{complete analytic function obtained from  $ ( f , G ) $: }
\begin{multline}
 \mathcal{F} [ ( f , G ) ] \; := \; \{ [ g]_{b} \, : \, \exists a
 \in G , \exists \gamma : [ 0 , 1 ] \mapsto \mathbb{C} \; : \;
\gamma(0) = a \, \wedge \, \gamma(1) = b \; path \; : \\
 [ g]_{b} \text{ is the analytic continuation of $ [ f]_{a} $ along $ \gamma
$ } \}
\end{multline}

Given $ \mathcal{F} $:
\begin{definition}
\end{definition}
\emph{$ \mathcal{F} $ is a complete analytic function:}
\begin{equation}
    \exists  \, ( f , G ) \text{ function element } \; : \;
    \mathcal{F} = \mathcal{F} [ ( f , G ) ]
\end{equation}

Given $ G \subset \mathbb{C} $ open:
\begin{definition} \label{sheaf of germs of analytic functions}
\end{definition}
\emph{sheaf of germs of analytic functions on G:}
\begin{equation}
    \mathcal{S} ( G ) \; := \; \{ ( z , [ f]_{z}) \; : \; z \in G
    , f \text{ is analytic at z } \}
\end{equation}

\bigskip

\begin{remark}
\end{remark}
It may be proved (see  \cite{Mc-Lane-Moerdijk-92}) that the
definition \ref{sheaf of germs of analytic functions} is a
particular case of the definition \ref{def:sheaf of germs}.

\bigskip

\begin{definition}
\end{definition}
\emph{projection map of $ \mathcal{S} ( G ) $:}

the map $ \pi :  \mathcal{S} ( G )  \mapsto \mathbb{C} $:
\begin{equation}
   \pi [ ( z , [f]_{z} ) ] \; := \; z
\end{equation}

Then:

\begin{proposition}
\end{proposition}
\begin{equation}
  \mathcal{S}_{z} ( G ) \; = \; \pi^{-1} (z) \; \; \forall z \in G
\end{equation}

where $ \mathcal{S}_{z} ( G ) $ is the set of germs in z of $
\mathcal{S}_{z} ( G ) $ defined, for an arbitrary sheaf of germs,
by the definition \ref{def:set of the germs of a sheaf in a point
of the base space}.

\smallskip

Given $ D \subset G $ open:
\begin{definition}
\end{definition}
\begin{equation}
    N( f , D ) \; := \; \{ ( z , [f]_{z} ) \, : \, z \in D \}
\end{equation}

Given $ ( a , [ f]_{a} ) \in \mathcal{S} (G) $:
\begin{definition}
\end{definition}
\begin{equation}
    \mathcal{N}_{ ( a , [ f]_{a} )} \; := \; \{ N( g , B ) \, : \, a \in B \wedge [g]_{a} = [ f]_{a} \}
\end{equation}

Then:
\begin{proposition}
\end{proposition}
\begin{enumerate}
    \item $ \{ \mathcal{N}_{ ( a , [ f]_{a} )} \, : \,  ( a , [ f]_{a} ) \in \mathcal{S}(G)
    \} $ is a neighborhood system on $  \mathcal{S}(G) $
    \item the induced topology  is Hausdorff
    \item the projection map $ \pi $ of $ \mathcal{S}_{z} ( G ) $
    is continuous with respect to such a topology
\end{enumerate}

\smallskip

Given a complete analytic function $ \mathcal{F} $ we have now all
the required ingredients to introduce the following:
\begin{definition} \label{def:Riemann surface of a complete analytic function}
\end{definition}
\emph{Riemann surface of $ \mathcal{F} $:}
\begin{equation}
    \Sigma_{\mathcal{F}} \; := \; \{ ( z, [ g]_{z} ) : [ g]_{z} \in \mathcal{F} \}
\end{equation}

\bigskip

\begin{remark}
\end{remark}
The definition \ref{def:Riemann surface of a complete analytic
function} allows to show that a complete analytic function is
indeed a function in the ordinary meaning of such a term: it may
be seen as the  map $ \mathcal{F} : \Sigma_{\mathcal{F}} \mapsto
\mathbb{C}$:
\begin{equation}
    \mathcal{F} [ ( z, [f]_{z} ) ] \; := \; f(z)
\end{equation}

\bigskip

Let us now introduce the second meaning of the locution "Riemann
surface":
\begin{definition} \label{def:Riemann surface}
\end{definition}
\emph{Riemann surface:}
\begin{center}
a one-complex-dimensional connected complex analytic manifold
\end{center}

The link existing between the definition \ref{def:Riemann surface
of a complete analytic function} and the definition
\ref{def:Riemann surface} is given by the following:

\begin{theorem}
\end{theorem}
\emph{Link between the two meanings of the locution "Riemann
surface":}
\begin{enumerate}
    \item the Riemann surface $ \Sigma_{f} $ of a complete
    analytic function f (according to the definition \ref{def:Riemann surface of a complete analytic
function}) is a Riemann surface (according to the definition
\ref{def:Riemann surface})
    \item for every  Riemann surface $ \Sigma $ (according to the
    definition \ref{def:Riemann surface}) satisfying suitable regularity conditions, there exists a complete
    analytic function f such that $ \Sigma $ is the Riemann surface of f (according to the definition \ref{def:Riemann surface of a complete analytic
function}), i.e such that:
\begin{equation}
    \Sigma \; = \; \Sigma_{f}
\end{equation}
\end{enumerate}

\bigskip

\newpage
\section{Notation}
\begin{center}
\begin{tabular}{|c|c|}
  \hline
  $ \wedge $ &  and (logical conjunction) \\
  $ \vee $ &   or (logical disjunction) \\
  $ \neg $ & not (logical negation) \\
  $ \vec{\nabla} f $ & gradient of the scalar field f \\
  $  \vec{\nabla} \wedge \vec{f}$ & rotor of the vector field $ \vec{f} $ \\
  $ \vec{\nabla} \cdot  \vec{f} $ & divergence of the vector field $ \vec{f} $  \\
  $ \alpha^{\sharp} $ & vector field associated to the 1-form $
  \alpha $ \\
  $ \vec{f}^{\flat} $ & 1-form  associated to the vector field $ \vec{f} $ \\
  $ \pi_{n} (M) $ & $ n^{th} $ homotopy group of M \\
  $ C_{n}(M) $ & $ n^{th} $ chain group of M \\
  $ Z_{n}(M) $ & $ n^{th} $ cycle group of M  \\
  $ B_{n}(M) $ &  $ n^{th} $ boundary group of M  \\
  $ H_{n}(M) $ & $ n^{th} $ homology group of M \\
  $ \Omega^{n}(M) $ & set of the n-forms over M \\
  $ \alpha \wedge \beta $ & exterior product of the form $ \alpha $ with respect to the form $ \beta $ \\
  $ d \omega $ & exterior derivative of the  form $
  \omega $ \\
  $ i_{X} \omega $ & interior product of the form $ \omega $ with
  respect to the vector field X \\
  $ C^{n}(M) $ & $ n^{th} $ cochain group of M \\
  $ Z^{n}(M) $ & $ n^{th} $ cocycle group of M  \\
  $ B^{n}(M) $ &  $ n^{th} $ coboundary group of M  \\
  $ H^{n}(M) $ & $ n^{th} $ de Rham cohomology group of M \\
  $[ G , G ] $ & commutator subgroup of the group G \\
  $ P( M, G) $  & principal bundle with base space M, total space
  P and structure group G \\
  $ T M $ & tangent bundle of M \\
  $ T^{\star} M $ & cotangent bundle of M \\
  $ \omega_{can} $ & canonical symplectic form of a cotangent
  bundle \\
  $ \Gamma ( T^{(r,s)} M ) $ & set of the global sections of the (r,s) tensor
  bundle over M \\
  $ \Gamma ( U, E ) $ & set of the local sections over U of the
  fibre bundle E \\
  $ \tilde{M} $ & universal covering space of M \\
  $ f_{\star} $ & differential map of f \\
  $ f^{\star} $ & pullback of f \\
  $ \star $ & Hodge Star operator associated to a riemannian
  metric \\
  $ \mathbb{C}^{\times} $ & punctured complex plane \\
 $ Ref(f, z) $ & residue of the function f in z \\
 $ \mathcal{S}(U)$ & group of sections of the sheaf $ \mathcal{S}
 $ over U \\
 $ \mathcal{S}_{x} $ & set of the germs of the sheaf $ \mathcal{S}
 $ in x \\
$ \Sigma_{f} $ & Riemann surface of the complete analytic function
f \\ \hline
\end{tabular}
\end{center}


\newpage
\begin{thebibliography}{10}

\bibitem{Nakahara-03}
M.~Nakahara.
\newblock {\em Geometry, Topology and Physics}.
\newblock Institute of Physics Publishing, Bristol and Philadelphia, 2003.

\bibitem{Arnold-89}
V.I. Arnold.
\newblock {\em Mathematical Methods of Classical Mechanics}.
\newblock Springer, Berlin, 1989.

\bibitem{Goldstein-Poole-Safko-02}
H.~Goldstein C. Poole~J. Safko.
\newblock {\em Classical Mechanics}.
\newblock Addison Wesley, 2002.

\bibitem{Feynman-63a}
R.~Feynman.
\newblock {\em The Feynman Lectures on Physics. Vol.1, Part 1. Mainly
  Mechanics, Radiation and Heat}.
\newblock Addison Wesley, 1963.

\bibitem{Abraham-Marsden-78}
R.~Abraham~J.E. Marsden.
\newblock {\em Foundations of Mechanics}.
\newblock Addison-Wesley, 1978.

\bibitem{Marsden-Ratiu-99}
J.E. Marsden~T.S. Ratiu.
\newblock {\em Introduction to Mechanics and Symmetry}.
\newblock Springer, New York, 1999.

\bibitem{De-Azcarrega-Izquierdo-95}
J.~De Azcarrega~J.M. Izquierdo.
\newblock {\em Lie Groups, Lie algebras, cohomology and some applications in
  physics}.
\newblock Cambridge University Press, Cambridge, 1995.

\bibitem{Fomenko-95}
A.~T. Fomenko.
\newblock {\em Symplectic Geometry}.
\newblock Overseas Publishers Association, Amsterdam, 1995.

\bibitem{McDuff-Salamon-98}
D.~Mc Duff~D. Salamon.
\newblock {\em Introduction to Symplectic Topology}.
\newblock Oxford University Press, Oxford, 1998.

\bibitem{Arnold-Givental-01}
V.I. Arnold~A.B. Givental.
\newblock Symplectic {G}eometry.
\newblock In V.I. Arnold~S.P. Novikov, editor, {\em Dynamical Systems 4.
  Symplectic Geometry and its Applications}, pages 4--138. Springer,
  Berlin, 2001.

\bibitem{Arnold-2005}
V.I. Arnold.
\newblock {\em Arnold's Problems}.
\newblock Springer, Berlin, 2005.

\bibitem{Dubrovin-Novikov-Fomenko-92}
B.~A. Dubrovin S.P. Novikov~A.F. Fomenko.
\newblock {\em Modern Geometry. Methods and Applications. Vol. 3: Introduction
  to Homology Theory}.
\newblock Springer, 1992.

\bibitem{Nash-94}
J.C. Nash.
\newblock {\em Differential Topology and Quantum Field Theory}.
\newblock Academic Press, London, 1991.

\bibitem{Balachandran-Marmo-Skagerstam-Stern-83}
A.P. Balachandran G. Marmo B.S. Skagerstam~A. Stern.
\newblock {\em Gauge Symmetries and Fibre Bundles}.
\newblock Springer, Berlin, 1983.

\bibitem{Lerda-92}
A.~Lerda.
\newblock {\em Anyons}.
\newblock Springer, Berlin, 1992.

\bibitem{Schulman-81}
L.S. Schulman.
\newblock {\em Techniques and Applications of Path Integration}.
\newblock John Wiley and Sons, New York, 1981.

\bibitem{Rivers-87}
R.~J. Rivers.
\newblock {\em Path Integral Methods in Quantum Field Theory}.
\newblock Cambridge University Press, Cambridge, 1987.

\bibitem{Cartier-De-Witt-Morette-06}
P.~Cartier C.~De Witt-Morette.
\newblock {\em Functional Integration: Action and Symmetries}.
\newblock Cambridge University Press, Cambridge, 2006.

\bibitem{Balachandran-Marmo-Skagerstam-Stern-91}
A.P. Balachandran G. Marmo B.S. Skagerstam~A. Stern.
\newblock {\em Classical Topology and Quantum States}.
\newblock World Scientific, Singapore, 1991.

\bibitem{Morandi-92}
G.~Morandi.
\newblock {\em The Role of Topology in Classical and Quantum Physics}.
\newblock Springer, Berlin, 1992.

\bibitem{Strocchi-05b}
F.~Strocchi.
\newblock {\em An Introduction to the Mathematical Structure of Quantum
  Mechanics. A Short Course for Mathematicians}.
\newblock World Scientific, Singapore, 2005.

\bibitem{Ryder-96}
L.H. Ryder.
\newblock {\em Quantum Field Theory}.
\newblock Cambridge University Press, Cambridge, 1996.

\bibitem{Weinberg-96}
S.~Weinberg.
\newblock {\em The Quantum Theory of Fields, vol.2. Modern Applications}.
\newblock Cambridge University Press, Cambridge, 1996.

\bibitem{Wilczek-90}
F.~Wilczek.
\newblock {\em Fractional Statistics and Anyon Superconductivity}.
\newblock World Scientific, Singapore, 1990.

\bibitem{Shokurov-94}
V.V. Shokurov.
\newblock Riemann {S}urfaces and {A}lgebraic {C}urves.
\newblock In I.R. Shafarevich, editor, {\em Algebraic Geometry I. Algebraic
  Curves, Algebraic Manifolds and Schemes}, pages 1--166. Springer, Berlin,
  1994.

\bibitem{Milnor-06}
J.~Milnor.
\newblock {\em Dynamics in One Complex Variable}.
\newblock Princeton University Press, Princeton, 2006.

\bibitem{Conway-78}
J.B. Conway.
\newblock {\em Functions of One Complex Variables I.}
\newblock Springer, New York, 1978.

\bibitem{Conway-95}
J.B. Conway.
\newblock {\em Functions of One Complex Variables II.}
\newblock Springer, New York, 1995.

\bibitem{Farkas-Kra-92}
H.M.~Farkas I.Kra.
\newblock {\em Riemann Surfaces}.
\newblock Springer, New York, 1992.

\bibitem{Mc-Lane-Moerdijk-92}
S.~Mac Lane~I. Moerdijk.
\newblock {\em Sheaves in Geometry and Logic. A First Introduction to Topos
  Theory}.
\newblock Springer, New York, 1992.

\bibitem{Isham-99}
C.J. Isham.
\newblock {\em Modern Differential Geometry for Physicists}.
\newblock World Scientific, Singapore, 1999.

\end{thebibliography}
\end{document}